\documentclass[twocolumn ,prl,longbibliography,a4paper,superscriptaddress,tightenlines,noeprint,10pt]{revtex4-2}
\RequirePackage{silence}
\RequirePackage{float}
\setcitestyle{compressed}
\usepackage[utf8]{inputenc}
\usepackage[USenglish]{babel}
\usepackage[T1]{fontenc}
\usepackage{lmodern}
\usepackage{array,multirow}
\usepackage{algorithm}
\usepackage{algorithmic}
\usepackage{graphicx,epsf}
\usepackage[caption=false]{subfig}
\usepackage[toc,page,header]{appendix}
\usepackage{minitoc}
\usepackage{tikz}
\usetikzlibrary{quantikz}
\usepackage[bookmarksnumbered,hypertexnames=false,colorlinks,colorlinks=true,linkcolor=blue,urlcolor=black,citecolor=blue,anchorcolor=green]{hyperref}
\usepackage{bbm,braket,microtype,mathrsfs,amsmath,amssymb,color,amsthm,mathtools,fullpage,enumitem,ae,aecompl,bm,thm-restate}
\usepackage{tikz}
\usepackage{lipsum}
\usepackage{caption}
\usepackage{color}
\usepackage{multirow}
\usepackage[capitalize]{cleveref}
\usepackage{silence}
\usepackage{newpxtext}
\usepackage{mathtools}

\allowdisplaybreaks
\newcommand{\be}{\begin{equation}}
\newcommand{\ee}{\end{equation}}
\newcommand{\ba}{\begin{eqnarray}}
\newcommand{\ea}{\end{eqnarray}}
\DeclareMathOperator{\tr}{tr}

\newcommand{\ignore}[1]{}
\newcommand{\st}[1]{\ket{#1}\bra{#1}}

\newcommand{\parent}[1]{\left( {#1} \right)}
\newcommand{\aver}[1]{ \left\langle  {#1}  \right\rangle }

\def\CC{{\rm\kern.24em \vrule width.04em height1.46ex depth-.07ex
   \kern-.29em C}}
\def\P{{\rm I\kern-.25em P}}
\def\RR{{\rm
        \vrule width.04em height1.58ex depth-.0ex
        \kern-.04em R}}
\def\bbbone{{\mathchoice {\rm 1\mskip-4mu l} {\rm 1\mskip-4mu l}
{\rm 1\mskip-4.5mu l} {\rm 1\mskip-5mu l}}}
\def\bbbc{{\mathchoice {\setbox0=\hbox{$\displaystyle\rm C$}\hbox{\hbox
to0pt{\kern0.4\wd0\vrule height0.9\ht0\hss}\box0}}
{\setbox0=\hbox{$\textstyle\rm C$}\hbox{\hbox
to0pt{\kern0.4\wd0\vrule height0.9\ht0\hss}\box0}}
{\setbox0=\hbox{$\scriptstyle\rm C$}\hbox{\hbox
to0pt{\kern0.4\wd0\vrule height0.9\ht0\hss}\box0}}
{\setbox0=\hbox{$\scriptscriptstyle\rm C$}\hbox{\hbox
to0pt{\kern0.4\wd0\vrule height0.9\ht0\hss}\box0}}}}
\def\bbbz{{\mathchoice {\hbox{$\sf\textstyle Z\kern-0.4em Z$}}
{\hbox{$\sf\textstyle Z\kern-0.4em Z$}}
{\hbox{$\sf\scriptstyle Z\kern-0.3em Z$}}
{\hbox{$\sf\scriptscriptstyle Z\kern-0.2em Z$}}}}

\newlength{\fighskip} \fighskip=2pt
\newlength{\figvskip} \figvskip=1pt
\newcommand*{\figbox}[2]{{
\def\figscale{#1}
\def\arraystretch{0.8}
\arraycolsep=0pt
\begin{array}{c}
\vbox{\vskip\figscale\figvskip
 \hbox{\hskip\figscale\fighskip
   \includegraphics[scale=\figscale]{#2}}}
\end{array}}}

\usepackage{hyperref}

\begin{document}
\setcounter{secnumdepth}{3}
\title{Retrieving information from a black hole using quantum
machine learning.}

\author{Lorenzo Leone}\email{Lorenzo.Leone001@umb.edu}
\affiliation{Physics Department,  University of Massachusetts Boston,  02125, USA}
\author{Salvatore F.E. Oliviero}
\affiliation{Physics Department,  University of Massachusetts Boston,  02125, USA}
\author{Stefano Piemontese}
\affiliation{Physics Department,  University of Massachusetts Boston,  02125, USA}

\author{Sarah True}
\affiliation{Physics Department,  University of Massachusetts Boston,  02125, USA}

\author{Alioscia Hamma}
\affiliation{Physics Department,  University of Massachusetts Boston,  02125, USA}
\affiliation{Dipartimento di Fisica Ettore Pancini, Universit\`a degli Studi di Napoli Federico II,
Via Cinthia 80126,  Napoli, Italy}


\begin{abstract}\noindent
In a seminal paper~\cite{hayden2007black}, Hayden and Preskill showed that information can be retrieved from a black hole that is sufficiently scrambling, assuming that the retriever has perfect control of the emitted Hawking radiation and perfect knowledge of the internal dynamics of the black hole. In this paper, we show that for $t-$doped Clifford black holes - that is, black holes modeled by random Clifford circuits doped with an amount $t$ of non-Clifford resources - an information retrieval decoder can be learned with fidelity scaling as $\exp(-\alpha t)$ using quantum machine learning while having access only to out-of-time-order correlation functions. We show that the crossover between learnability and non-learnability is driven by the amount of non-stabilizerness present in the black hole and sketch a different approach to quantum complexity.
\end{abstract}
 \maketitle

\section{Introduction}
The onset of chaos is at the root of the explanation of thermalization in closed quantum systems~\cite{lloyd1988black,srednicki1994chaos,popescu2006thermal,popescu2006entanglement,rigol2010chaos}. Although the precise definition of quantum chaos remains elusive, one can usefully refer to it as a bundle of features comprising information scrambling~\cite{hosur2016chaos,ding2016conditional,brown2012scrambling,yan2020information,touil2020quantum}, a complex pattern of entanglement~\cite{chamon2014emergent,yang2017entanglement,liu2018generalized,liu2018entanglement, true2022transitions}, universal behavior of out-of-time-order correlation functions (OTOCs)~\cite{larkin1969quasiclassical,kitaev2014hidden,roberts2017chaos,Luitz2017information,Hashimoto2017otoc,halpern2017otoc,garciamata2018chaosotoc,Rakovsky2018otoc,lin2018out,santos2019otoc,fortes, leone2020isospectral} and quantum dynamics that can be efficiently modeled by random unitary operators~\cite{brown2010random,brown2010convergence,roberts2017chaos,harrow2018approximate,gharibyan2018onset}.  
Information scrambling is the quantum analog of the butterfly effect, signaling that local disturbances are spread through operator growth~\cite{roberts2015chaos,cotler2018out,nahum2018operator}. 

In the context of black hole physics, one wonders whether the information is destroyed by a black hole or can be recovered from Hawking radiation as the black hole evaporates. It is commonly assumed that a black hole thermalizes information quickly~\cite{lloyd1988black,srednicki1994chaos,rigol2016thermalization} and that information is also rapidly spread across every part of the system, making the black hole a fast scrambler~\cite{sekino2008fast,lashkari2013towards,shenker2014multi,maldacena2016bound}. Moreover, one assumes evolution to be described by a unitary operator $U$. Under these assumptions, Hayden and Preskill showed that the Hawking radiation releases information very quickly and that one can therefore recover any information initially dumped into the black hole with just a slight overhead of information read-out from the subsequent Hawking radiation~\cite{hayden2007black}. 
This remarkable result hinges on the fact that it is the very scrambling dynamics of the black hole that allows information to be transferred to the Hawking radiation. Yoshida and Kitaev showed how this information can be retrieved by an observer with perfect knowledge of both the initial state of the black hole {\em and} its unitary dynamics $U$~\cite{Yoshida2017efficient}.

Obtaining perfect knowledge of a black hole’s internal structure and dynamics sounds like an impossibly daunting task. Could we perhaps learn $U$ by feeding the Hawking radiation into a suitable quantum machine learning (QML) algorithm~\cite{Biamonte2017qml,Ciliberto2018qml,Dunjko2018qml,Lloyd2018qml,osborne2020nofree,Lloyd2020qml,Lloyd2020qml2,McClean2016variational,Romero2017variational,LaRose2019variational,Arrasmith2019variational,Cerezo2020variationalquantum,Sharma2020variational,Prieto2019variational,cerezo2020variational,Commeau2020variational,Li2017variational,Yuan2019theoryofvariational,sharma2020machlearn,Volkoff2021gradients,Endo2020MC}? Such an approach seems promising at first glance, but quickly becomes futile in practice; in order for the recovery algorithm to work, the black hole must be scrambling, but the supposed complexity of scrambling dynamics hinders our ability to learn about its details. Indeed, extensive analysis~\cite{Holmes2021barren,McClean2018barrenp,Cerezo2021barren,Wang2021barren,Cerezo2021barren2,Arrasmith2021effectofbarren,Grant2019initialization,kiani2022barren,garcia2022barren,leone2022HEA} of the barren plateau problem—i.e. system size-exponential vanishing of cost function gradients in variational quantum algorithms—has shown that no QML training protocol can effectively learn $U$ if it is scrambling.

And yet, even so, things are not hopeless. We cannot learn $U$, but we can train a quantum circuit to decode it by learning a model unitary $V$ that is good enough to unscramble the Hawking radiation, recover the information tossed into the black hole, and perform teleportation. Despite being very different from the original scrambler $U$, this so-called mocking black hole $V$ is optimized to perform the desired tasks.

In this paper, we show that if a black hole is modeled by a unitary $U_t$ consisting of an element of the Clifford group with a doping $t$ of non-Clifford resources, one can use a cost function directly obtained from the OTOCs to learn a mocking $V$ capable of recovering information from the black hole with fidelity $\mathcal{F}(V)\sim\exp(-\alpha t)$. This is possible because, while the Clifford group is a good scrambler~\cite{scott2008optimizing,zhu2017multiqubit,zhu2016clifford}, it does not produce a complex pattern of entanglement across the subsystems~\cite{leone2021quantum,oliviero2021transitions}. We also show that, with the same technique, one can teleport a state with zero initial knowledge of black hole dynamics.

\onecolumngrid
\begin{center}
\begin{figure}[H]
\centering
\includegraphics[width=\linewidth]{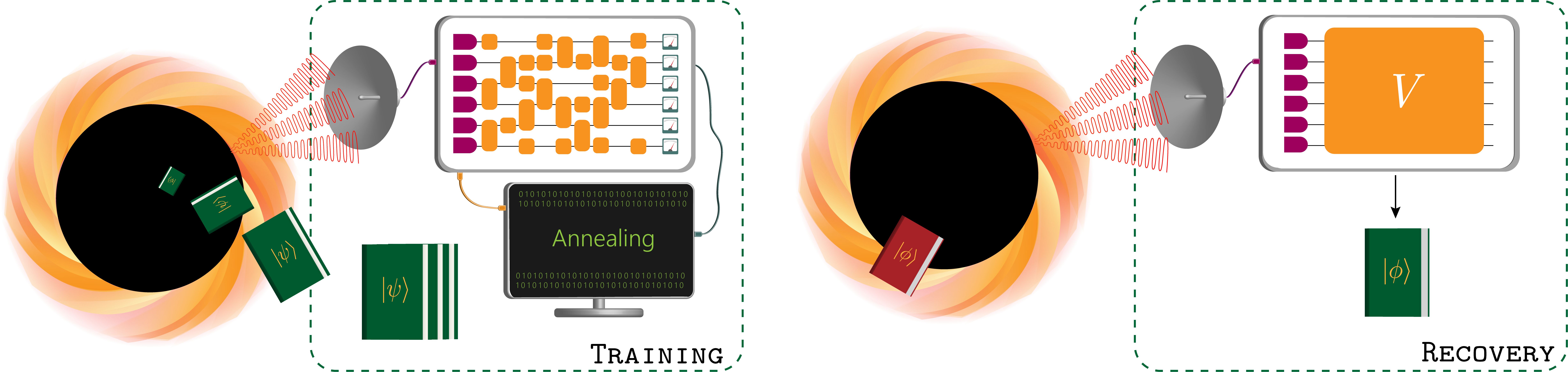}   
\caption{Quantum machine learning protocol for learning the mocking unitary $V$ capable of performing information recovery from Hawking radiation. Using a cost function given by OTOCs, the model learns to recover a set of known training states (the green journals) that are tossed into the black hole. This model unitary $V$ is then used to decode the Hawking radiation emitted after tossing in a new, unknown journal and retrieve the information contained within it.}
\label{figmain}
\end{figure}
\end{center}
\twocolumngrid
\noindent

\section{Notation and techniques}\label{tools}
Let us introduce some notations and tools that are used throughout the paper. Consider the Hilbert space of $n$ qubits $\mathcal{H}$, partitioned as $n=n_A+n_B=n_C+n_D$. Let $d_{\Lambda}\equiv 2^{n_{\Lambda}}$ with $\Lambda\in\{A,B,C,D\}$. Denote $\mathcal{P}_{n_{\Lambda}}$ the Pauli group on $n_{\Lambda}$ qubits, being $\Lambda\in\{A,B,C,D\}$. Let $P_{\Lambda}\in\mathcal{P}_{\Lambda}$ be local Pauli operators acting non-identically only on $\Lambda$, where we omit the identity on the complement of $\Lambda$. Let $O$ be a operator on $\mathcal{H}$ and let $P_{\Lambda}OP_{\Lambda}$ be the twirling of $O$ with $P_{\Lambda}$; denoting the average over the Pauli group $\mathcal{P}_{n_{\Lambda}}$ as $\aver{P_{\Lambda}OP_{\Lambda}}_{\mathcal{P}_{n_\Lambda}}\equiv d_{\Lambda}^{-2}\sum_{P_\Lambda\in\mathcal{P}_{n_\Lambda}}P_{\Lambda}OP_{\Lambda}$, we have
\be
\aver{P_{\Lambda}OP_{\Lambda}}_{P_\Lambda}=\frac{\tr_\Lambda(O)}{d_{\Lambda}}\,,
\label{averagepauligroup}
\ee
where $\tr_{\Lambda}(\cdot)$ denotes the partial trace on $\Lambda$. Consider a Hilbert space $\mathcal{H}_{R}$ of $n_A$ auxiliary qubits, then the (Einstein-Podolsky-Rosen) EPR pair between $A$ and $R$ is denoted as $\ket{RA}$, living on $\mathcal{H}_R\otimes \mathcal{H}_A$, and is defined as
\be
\ket{RA}\equiv \frac{1}{\sqrt{d_{A}}}\sum_{i=1}^{d_A}\ket{i}_R\otimes\ket{i}_A\,.
\ee
Denote $\Pi_{\Lambda\Lambda'}\equiv\st{\Lambda\Lambda^{\prime}}$ the projector onto the EPR pair $\ket{\Lambda\Lambda^{\prime}}$. In what follows, we adopt the following pictorial representation for EPR pairs
\be
\ket{RA}=\figbox{.3}{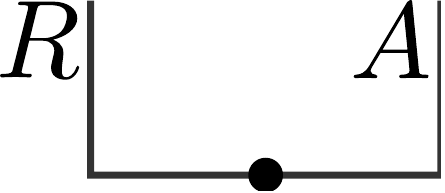}\,,
\label{eprrepresentation}
\ee
where the horizontal lines represent the EPR pair and the dot capture the normalization $d_{A}^{-1/2}$. Let $O_A$ a unitary operator on $\mathcal{H}_A$, then $O_A\ket{RA}=O_{R}^{T}\ket{RA}$, where $T$ is the transposition. Diagrammatically
\be
O_{R}^{T}\ket{RA}=\figbox{.25}{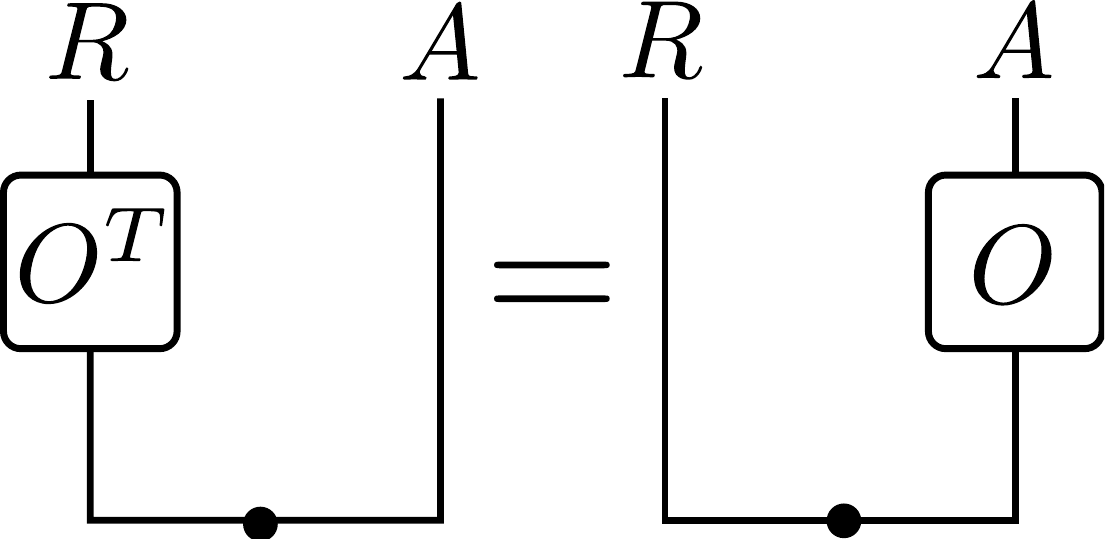}=O_A\ket{RA}\,.
\ee
Consider two EPR pairs $\ket{RA}\ket{BB^{\prime}}$, where $B^{\prime}$ is a $n_B$ qubits auxiliary system, and consider a unitary operator $O_{AB}$ acting on $A$ and $B$, as $O_{AB}\ket{AR}\ket{BB^{\prime}}$. The diagrammatic representation corresponding to the projection of $O_{AB}\ket{AR}\ket{BB^{\prime}}$ onto $\bra{RA}$ is
\be
\bra{RA}O_{AB}\ket{RA}\ket{BB^{\prime}}=\figbox{.25}{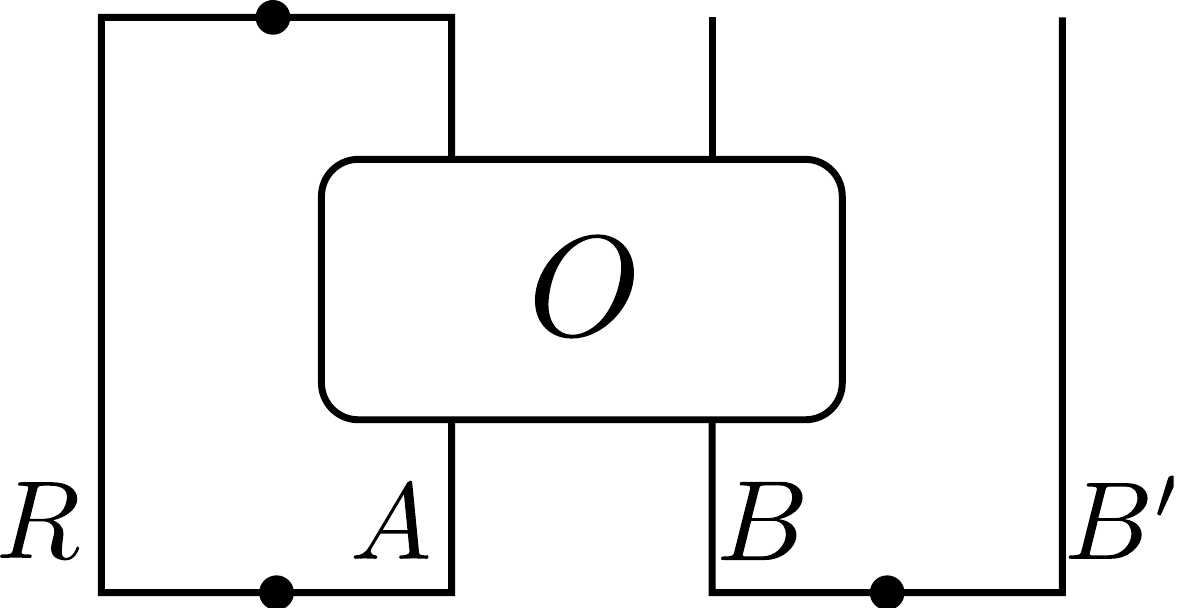}\,.
\ee
Noting that $\bra{RA}O_{AB}\ket{RA}\ket{BB^{\prime}}=d_{A}^{-1}(\tr_{A}O_{AB})\ket{BB^{\prime}}$, we can use the identity in Eq.~\eqref{averagepauligroup} to establish the following relation between diagrams
\be
\frac{1}{d_A^2}\sum_{P\in\mathcal{P}_{n_A}}\figbox{.2}{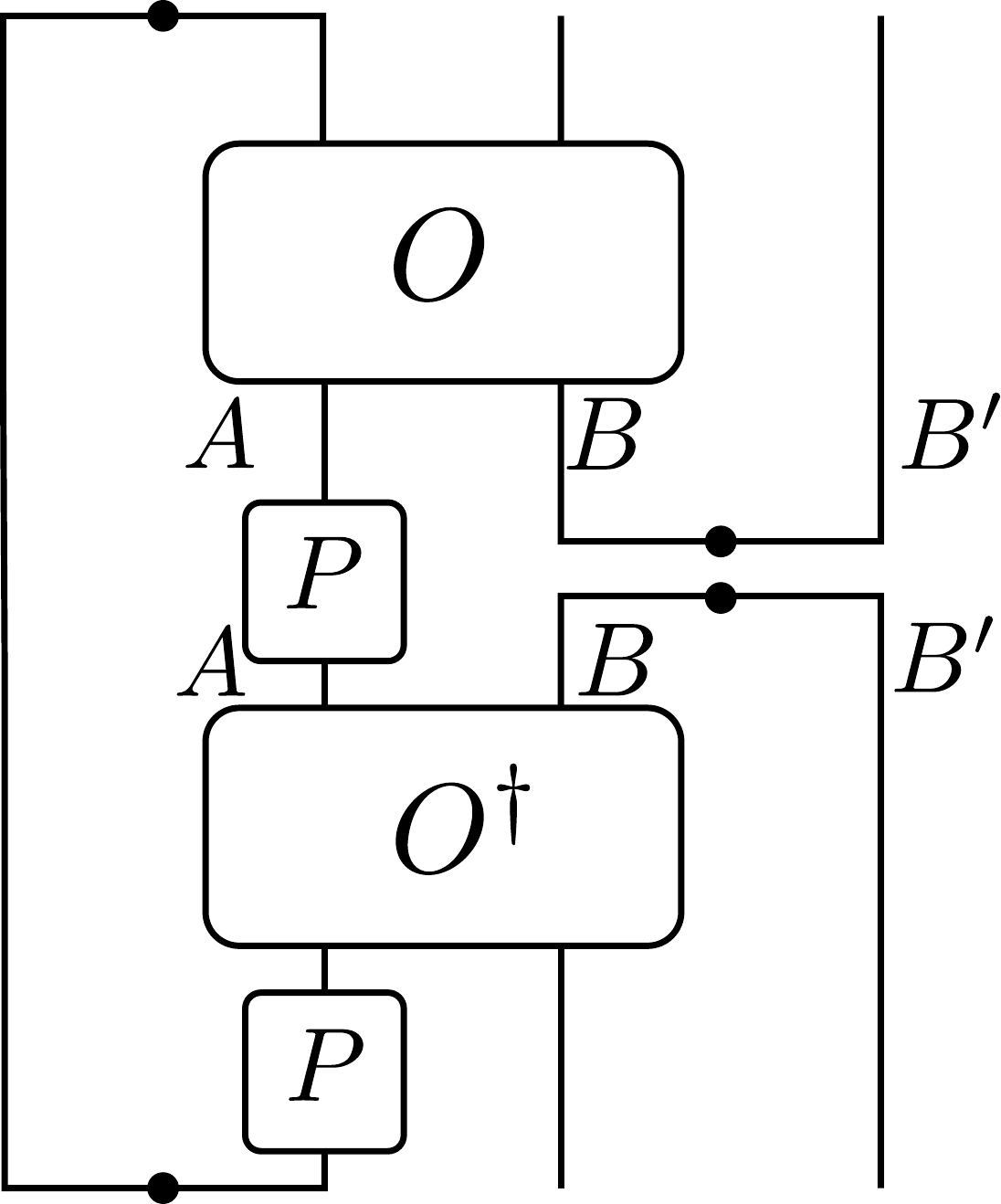}=\figbox{.2}{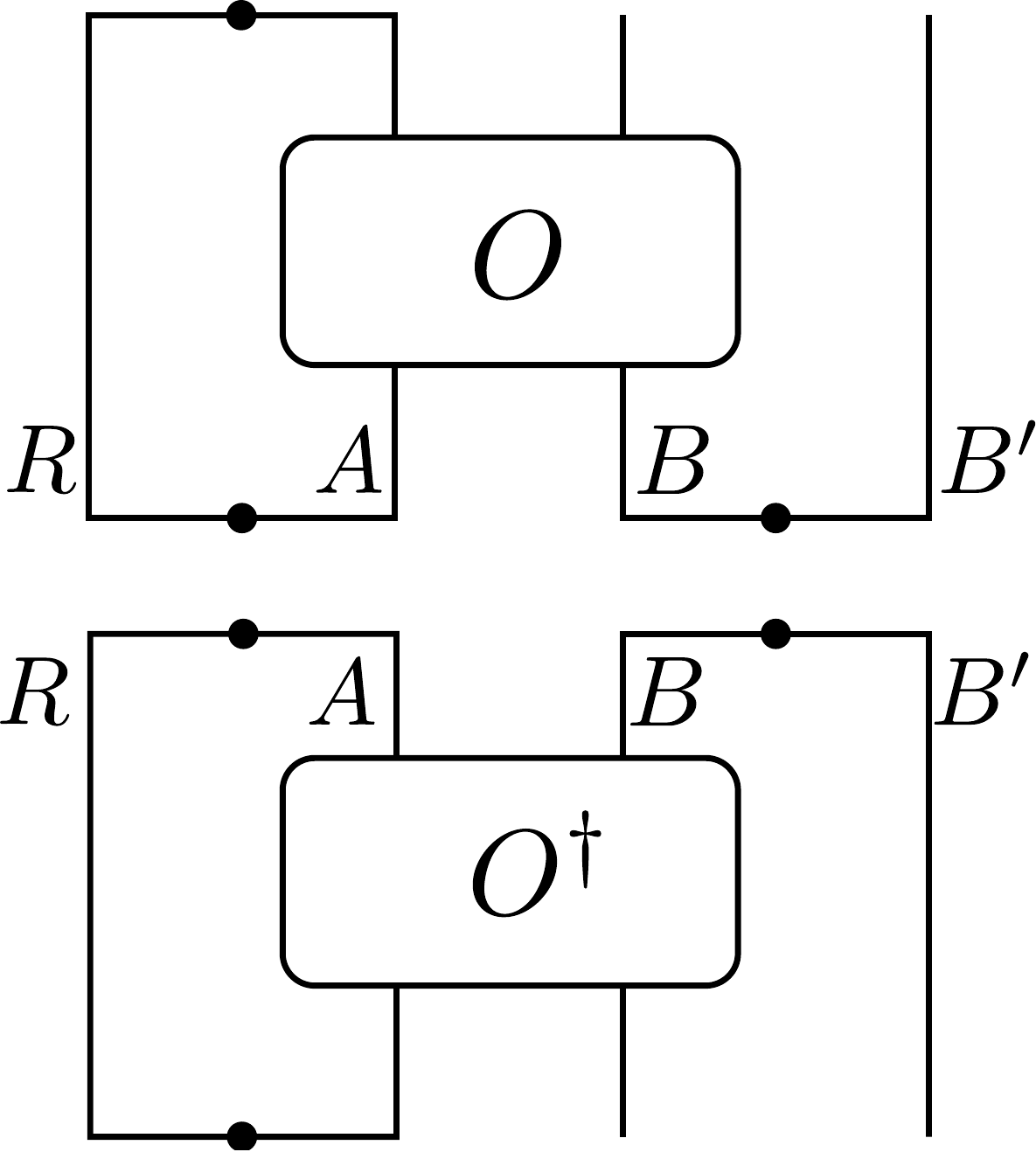}
\label{graphicpauliav}\,,
\ee
where we considered the diagram corresponding to the projector $d_{A}^{-2} (\tr_A O_{AB})\st{BB^{\prime}}(\tr_A O_{AB}^{\dag})$. The above relation will be useful to prove the main equations of the paper.

\section{Decoding a Black Hole}
Let us start by reviewing the decoding algorithm for information retrieval from a scrambling black hole~\cite{hayden2007black, Yoshida2017efficient}. 
After a long-time evaporation process, the state of the black hole $B$ is maximally entangled with the already emitted Hawking radiation $B^{\prime}$, which can be accessed by the observer Bob. This state is described by an EPR pair $\ket{BB'}$.  Alice possesses a quantum state $A$ maximally entangled with a reference state $R$, that is an EPR pair $\ket{RA}$. The maximal entanglement between $A$ and $R$ quantifies the information about $A$ possessed by $R$. This information is lost when the black hole destroys correlations between $A$ and $R$. The initial state of the whole system is thus described by $\ket{RA}\ket{BB^{\prime}}$.

Alice tosses her quantum state - depicted as a journal in Fig.~\ref{figmain} - into the black hole, which will soon emit new Hawking radiation $D$ and retain internal degrees of freedom labeled by $C$. So $A,B$ label the internal degrees of freedom of the black hole before Alice tosses her journal into it and $B^{\prime}$ labels the Hawking radiation already emitted at that point. After tossing the journal, the internal degrees of freedom of the black hole are described by a state on $C$ and the Hawking radiation is given by $D+B^{\prime}$.  

The black hole dynamics are modeled by a unitary $U=I_{RB'}\otimes U_{AB}$ which must be random enough to scramble information. After the evolution $U$, the state of the entire system is given by the pure state $\ket{\Psi}_{RB^{\prime}CD}=I_{RB'}\otimes U_{AB}\ket{RA}\ket{BB^{\prime}}$ in $RB^{\prime}CD$, which diagrammatically is given by
\be
\ket{\Psi}_{RB^{\prime}CD}=\figbox{.3}{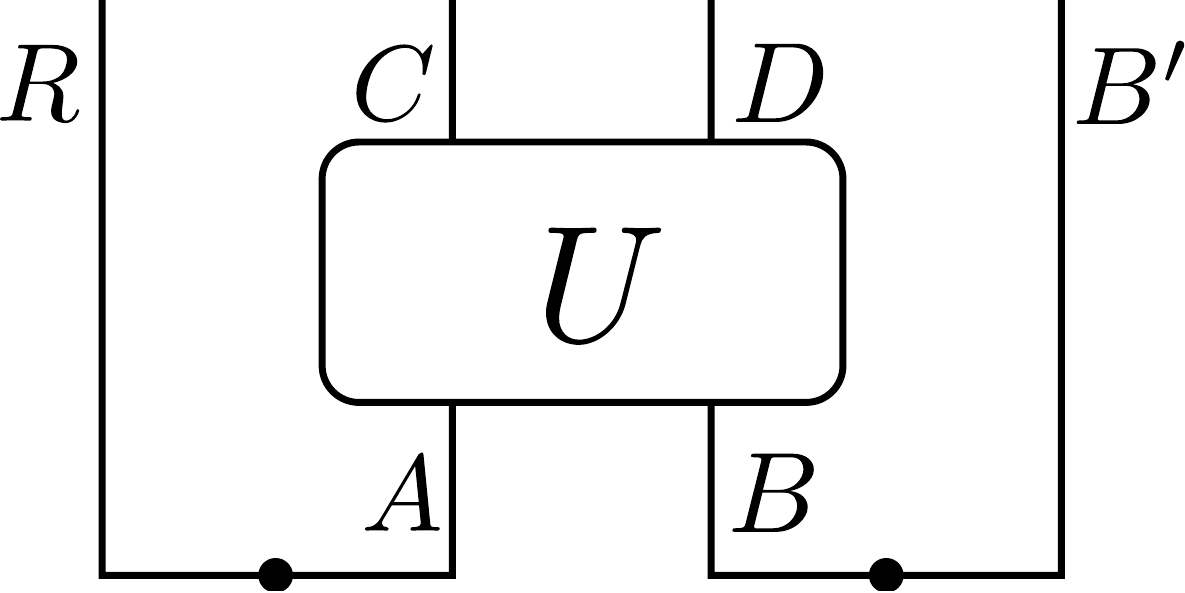}\,.
\ee
Notice that the number $n$ of qubits initially internal to the black hole obeys  $n\equiv n_A+n_B=n_C+n_D$, so the Hilbert space of the black hole interior is shrinking as more Hawking radiation is emitted. 


The information possessed by Alice in her journal would be recovered by Bob if he could process the Hawking radiation and end up with a state maximally entangled with another reference state $R'$ in his possession. In the case that Alice's state is a pure state $\ket{\psi_A}$, this would amount to teleporting $\ket{\psi_A}$ from $A$ to $R'$. 

As~\cite{hayden2007black} showed, the existence of such a recovery procedure is contingent on the unitary $U$ being scrambling enough. Because of unitarity, correlations can only be transferred, so if $U$ is scrambling enough to destroy correlations between $R$ and $C$, these must show up somewhere else and can be decoded by a suitable $V^*$. 
The information shared between the qubits of Alice $R$ and the internal degrees of freedom of the black hole $C$ is given by the mutual information $I(R|C):=S(\rho_{R})+S(\rho_{C})-S(\rho_{RC})$, where $S(\rho_{\Lambda})\equiv-\tr\rho_{\Lambda}\log\rho_{\Lambda}$ is the Von Neumann entropy, and $\rho_{\Lambda}:=\tr_{\bar{\Lambda}}\st{\Psi}$ (with $\bar{\Lambda}$ being the complement of $\Lambda$) are reduced density matrices for $\Lambda\in\{R,C,RC\}$.  One can easily see~\cite{hosur2016chaos} that for the state $\ket{\Psi}_{RB^{\prime}CD}$ one has $I(R|C)=n_A+n_C-S(\rho_{RC})$. Define the OTOC as $
\Omega(U):={d^{-1}}\aver{\tr(P_AP_{D}(U)P_AP_{D}(U))}_{P_A,P_{D}}$, where $P_{D}(U)\equiv U^{\dag}P_DU$; then, the mutual information can be related to the OTOCs $\Omega(U)$, by $S_{2}(\rho_{RC})=-\log \frac{d_{A}}{d_{C}}\Omega(U)$ proven in~\cite{hosur2016chaos}, as
\be
I(R|C)\le -\log d_{A}^{2}\Omega(U)\,.
\label{mutualinfo}
\ee
A unitary $U$ is said to be \textit{scrambling } if and only if $\Omega(U)\simeq {d_{A}^{-2}}+{d_{D}^{-2}}-{(d_{A}d_{D})^{-2}}$, where $\simeq$ means \textit{up to an order $d^{-2}$}. In the limit of $d_{A}\ll d_{D}$, the scrambling dynamics will therefore imply that $I(R|C) \simeq 0$, thus resulting in the destruction of correlations between $R$ and $C$. Thanks to the unitarity of $U$, all the information has been transferred to Bob; the mutual information $I(R|DB^{\prime})$ between $R$ and the qubit in Bob's possession $B^{\prime}$ together with the new emitted Hawking radiation $D$ is maximal, i.e.
\be
I(R|DB^{\prime})\simeq n_A\,.
\ee
Let us see how Bob is able to recover the information initially possessed by Alice. Bob possesses an EPR pair $A'R'$ and applies a unitary $V^*$ to the old Hawking radiation $B'$ and one-half of his pair $A'$. Then, by reading the additional Hawking radiation $D$, effectively entangling it in another EPR pair $DD'$, Bob projects by $\Pi_{DD^{\prime}}$ onto $DD'$ to obtain a final state
$\ket{\Psi_{out}}=\frac{1}{\sqrt{P_{out}}}\Pi_{DD^{\prime}}V^{*}_{B^{\prime}A^{\prime}}U_{AB}\!\ket{RA}\!\ket{BB^{\prime}}\!\ket{A^{\prime}R^{\prime}}$,
where $P_{out}$ is a normalization. Diagrammatically:
\be
\ket{\Psi_{out}}=\frac{1}{\sqrt{P_{out}}}\figbox{0.13}{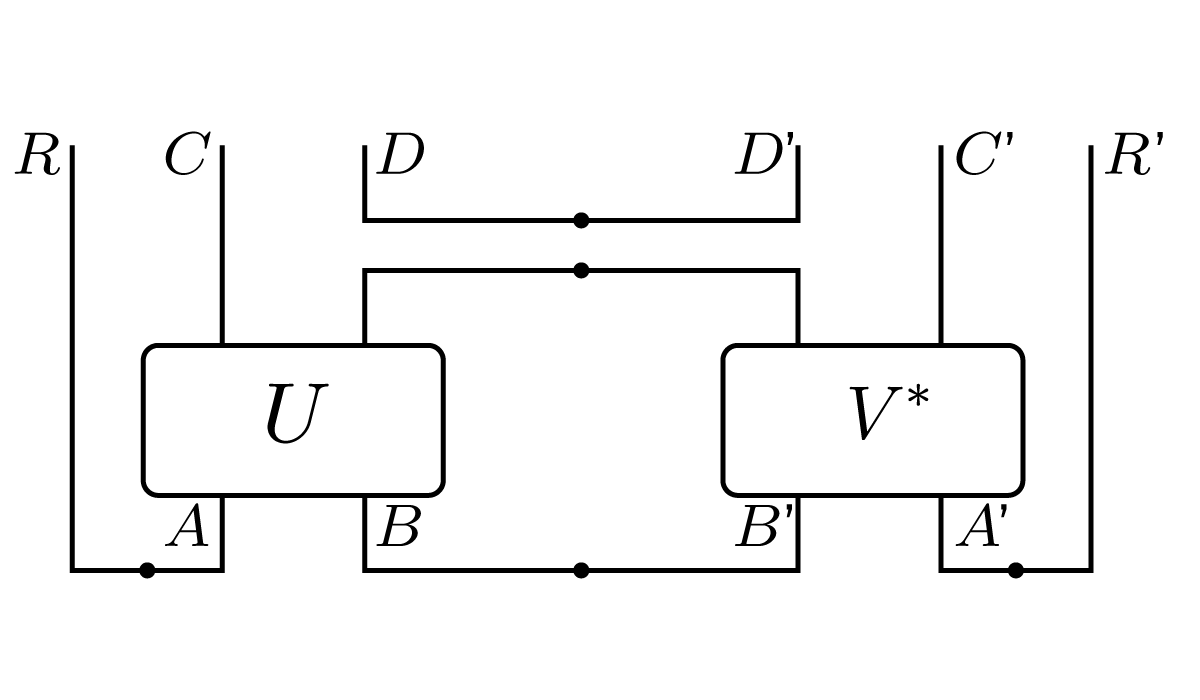}\,.
\label{scheme}
\ee
Following~\cite{Yoshida2017efficient,Yoshida2019disentangling,li2021hayden}, the normalization $P_{out}$ can be computed with the help of the diagrammatic formalism introduced in Sec.~\ref{tools} as:
\be
P_{out}\equiv \figbox{0.13}{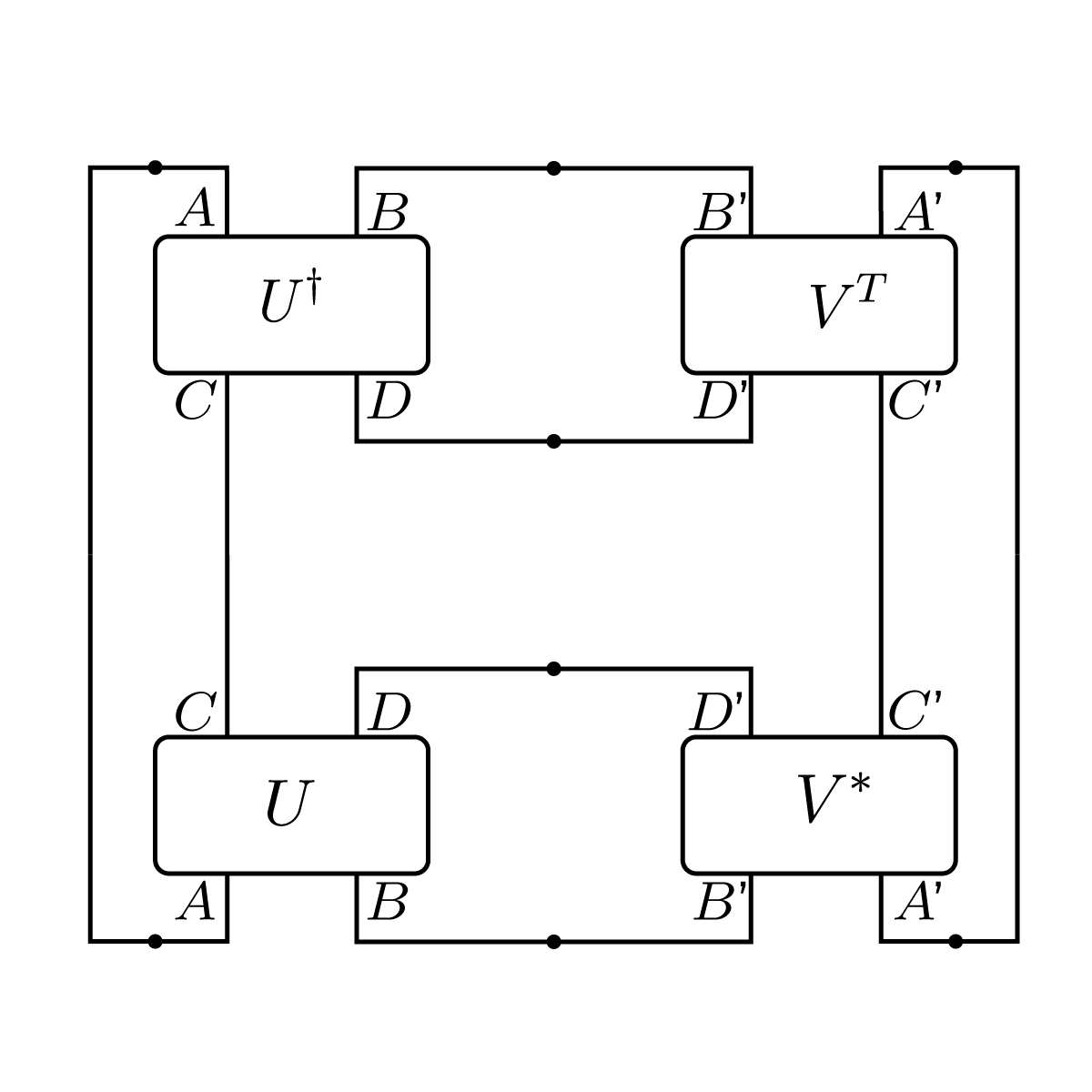}\,.
\ee
The above diagram is obtained from contracting the diagram corresponding to $\Pi_{DD^{\prime}}V^{*}_{B^{\prime}A^{\prime}}U_{AB}\!\ket{RA}\!\ket{BB^{\prime}}\!\ket{A^{\prime}R^{\prime}}$ with itself. By using Eq.~\eqref{averagepauligroup} and its diagrammatical representation in Eq.~\eqref{graphicpauliav}, one arrives to
\be
P_{out}=\frac{1}{d}\aver{\tr(P_{D}(U)P_{A}P_{D}(V)P_{A})}_{P_A,P_D}\,.
\label{1}
\ee
The recovery algorithm is successful if one obtains a maximally entangled pair between $R$ and $R'$ that is factorized from the rest. In this way, the information about $A$ originally in the hands of Alice has been successfully transferred to Bob~\cite{hayden2007black}. In other words, the final state must read $\ket{\Psi_{out}}\simeq \ket{RR^{\prime}}\otimes \ket{rest}_{CC^{\prime}}\otimes \ket{DD^{\prime}}$. Success of the algorithm can be established by computing the fidelity $\mathcal{F}(V):=\tr(\Psi_{out}\Pi_{RR^{\prime}})$, where $\Psi_{out}\equiv \st{\Psi_{out}}$; the product $\mathcal{F}(V)P_{out}$ can be easily computed using diagrammatical techniques as
\be
\mathcal{F}(V)P_{out}=\figbox{0.067}{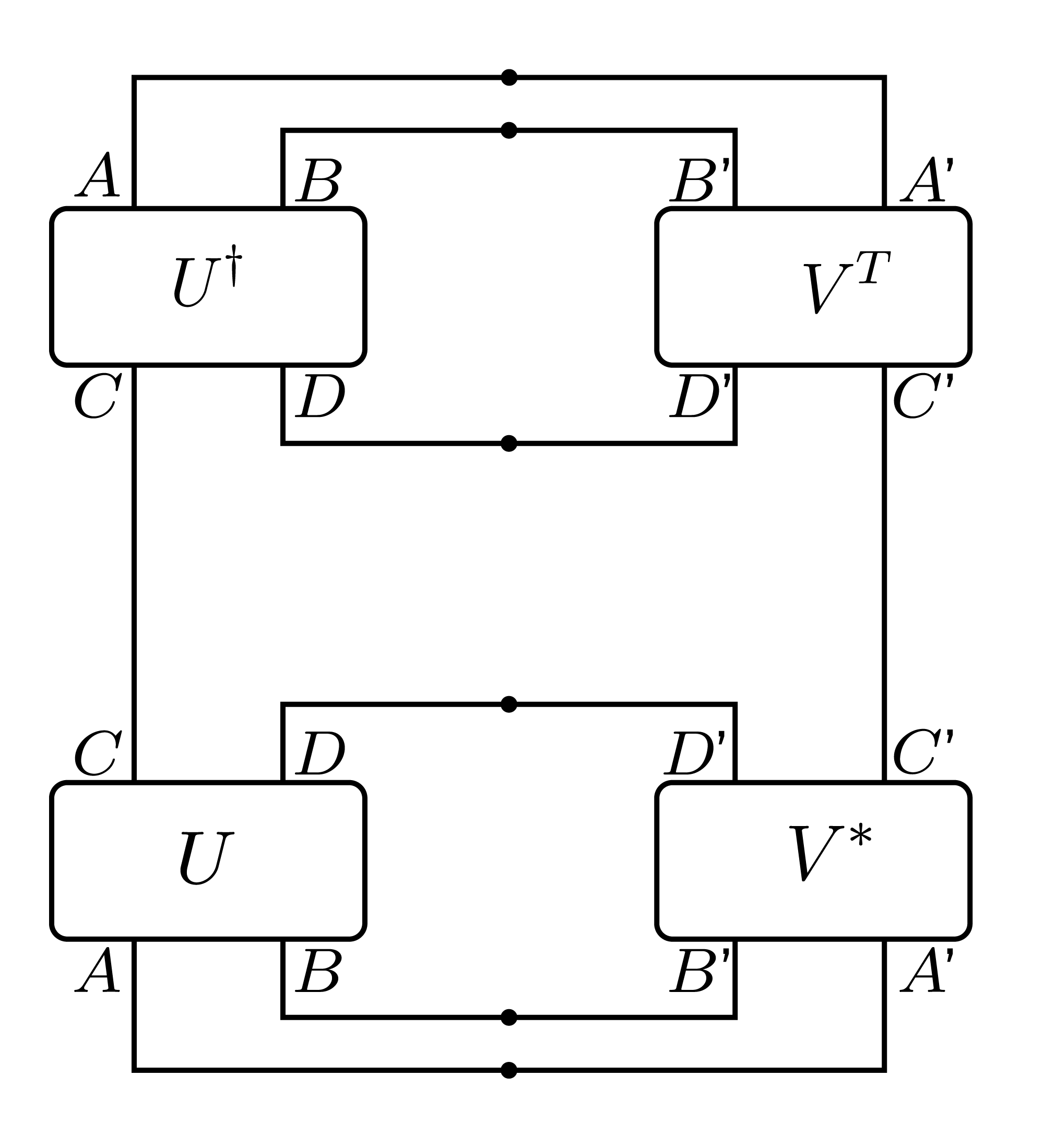}\,,
\ee
which is obtained by crunching the diagram in Eq.~\eqref{scheme} with the one corresponding to $\ket{RR^{\prime}}$ (see Eq.~\eqref{eprrepresentation}). Using Eq.~\eqref{graphicpauliav} one arrives to 
\be
\mathcal{F}(V)P_{out}=\frac{1}{d_A^2d}\aver{\tr{P_D(U)P_D(V)}}_{P_D}\,.
\label{2}
\ee
Putting together Eq.~\eqref{1} and Eq.~\eqref{2}, one derives the main equation of the paper, i.e. the recovery fidelity $\mathcal{F}(V)$ as a function of the scrambler $U$ and the decoder $V$:
\be
\mathcal{F}(V)=\frac{1}{d_{A}^{2}}\frac{\aver{\tr(P_{D}(U)P_{D}(V))}_{P_D}}{\aver{\tr(P_{D}(U)P_{A}P_{D}(V)P_{A})}_{P_A,P_D}}\,.
\label{fidelityunknown}
\ee
In the above expression, the role of $V$ is to {\em mock} the behavior of the black hole modeled by $U$. If $V=U$, the behavior of $U$ is obviously replicated perfectly. Indeed, the result in~\cite{Yoshida2017efficient} shows that in the ideal case $V=U$, the fidelity $\mathcal{F}(U)$ reads
$\mathcal{F}(U)=\frac{1}{d_{A}^{2}\Omega(U)}$,
and one can see that if the black hole is indeed scrambling, one obtains a fidelity  $\mathcal{F}(U)= 1-O(d_{A}^{2}/d_{D}^{2})$. One can easily see~\cite{Yoshida2019disentangling} that
$\tr(P_{D}(U)P_{A}P_{D}(V)P_{A})=\bra{\Psi_{out}^{P_A}}P_{D}\otimes P_{D^{\prime}}^{*}\ket{\Psi^{P_A}_{out}}$
where $\ket{\Psi_{out}^{P_A}(U,V)}:=(U_{AB}\otimes V^{*}_{B^{\prime}A^{\prime}})(P_{A}\otimes 1_{A^{\prime}RR^{\prime}BB^{\prime}})\ket{RA}\ket{BB^{\prime}}\ket{A^{\prime}R^{\prime}}$. Thus, $\mathcal{F}(V)$ is a quantity that can be recast as a ratio of expectation values of local observables (on $D$) with supports on accessible parts of the system, being $D$ the emitted Hawking radiation. Note that, if the qubits in $A$ are $n_A=O(1)$, the new emitted Hawking radiation $D$ that is collected by Bob must obey to $n_{D}=n_{A}+\log\epsilon^{-1/2}$ to feature a $\epsilon$-maximal mutual information $I(R|DB^{\prime})=n_A-\epsilon$, cf. Eq.~\eqref{mutualinfo}. Thus, collecting slightly more many qubits than the ones contained in $A$, Bob decodes the scrambled information by local $O(1)$ measurements.

\subsection{ Learning the Black Hole decoder} 
Can we learn the Black hole by observing its Hawking radiation by some quantum machine learning algorithm? The presence of barren plateaus~\cite{Holmes2021barren,McClean2018barrenp,Cerezo2021barren,Wang2021barren,Cerezo2021barren2,Arrasmith2021effectofbarren,Grant2019initialization,kiani2022barren} seems to forbid any kind of learning if the black hole is scrambling enough, which is also the condition that allows information retrieval. 
The main result of this paper is that - under suitable conditions - one can nevertheless learn a decoder $V$ that mocks the behavior of the black hole sufficiently enough to recover information. 

\begin{figure}[H]
\centering
\includegraphics[width=\linewidth]{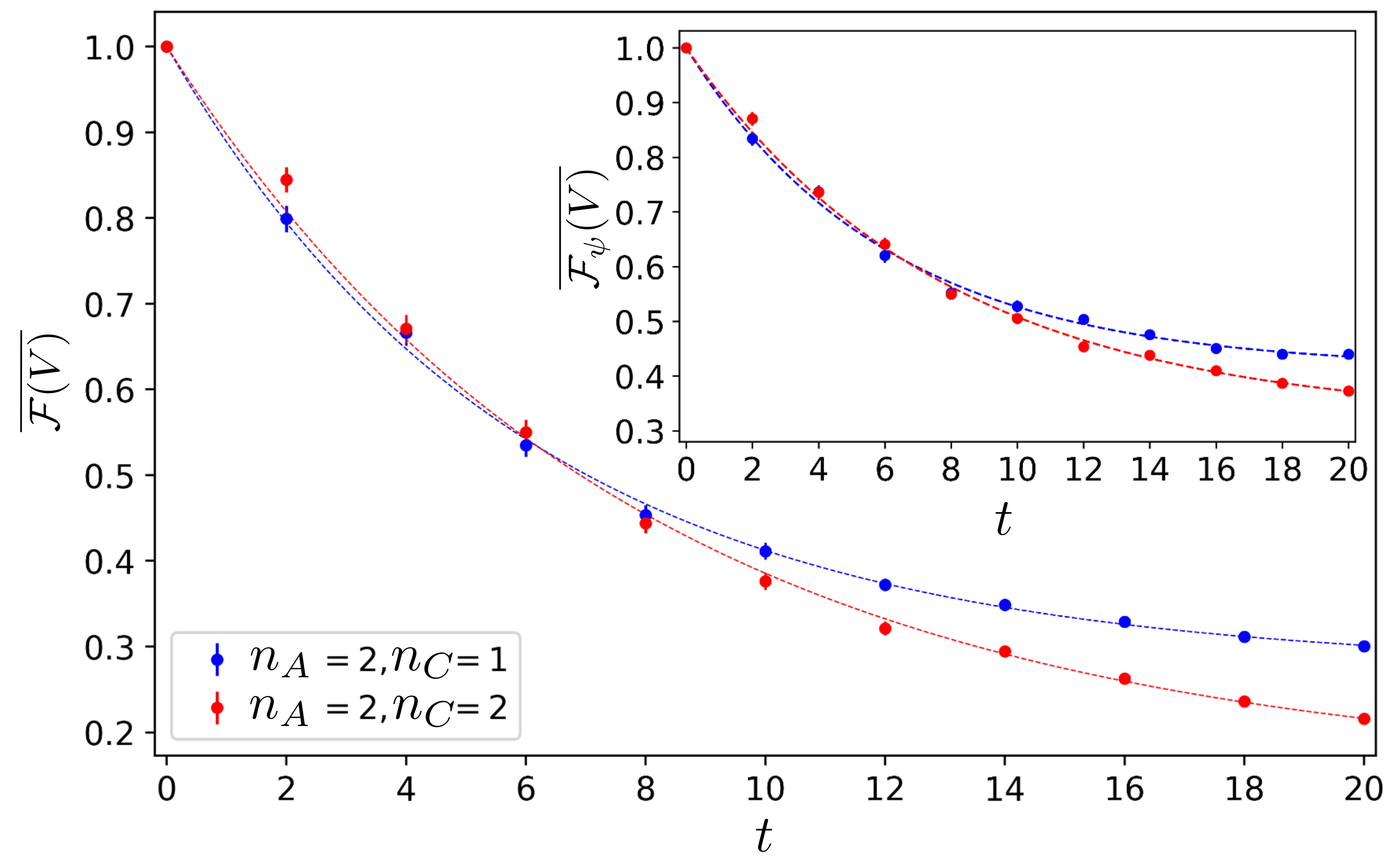}   
\caption{Average over $200$ realizations of the fidelity $\mathcal{F}(V)$, denoted as $\overline{\mathcal{F}(V)}$, at the end of the Metropolis Algorithm as a function of the doping $t$, fit to $a\,\mathrm{exp}(-\alpha t)+b$ where $\alpha=0.167$, $a=0.72$, $b=0.28$ for $n_C=1$, while $\alpha=0.129$, $a=0.85$, $b=0.15$ for $n_C=2$. Each realization is performed for a newly generated black hole unitary $U$. \textit{Inset:} Same for the teleportation fidelity $\mathcal{F}_{\psi}(V)$, with the same values for $\alpha$, see App~\ref{appb}.}
\label{fig1}
\end{figure}

Consider the case that the black hole is modeled by  a  $t-$doped random Clifford circuit $U\in\ \mathcal{C}_{t}:=\{C_{t-1}T_{t}C_{t-2}\cdots T_{2}C_{1}T_{1}C_{0}\,|\, C_{k}\in\mathcal{C}(2^n)\}$ with $T_{k}$ being single qubit $T-$gates applied on random qubits~\cite{leone2021quantum}. Such operators are scrambling enough~\cite{yoshida2021decoding, leone2021quantum} for every $t$. However, for $t=0$ - i.e. random Clifford operators - a learning protocol based on a Metropolis algorithm is possible~\cite{chamon2014emergent, shaffer2014irreversibility, zhou2020single}. Moreover, it has been recently shown that low $t-$doping random Clifford circuits can be efficiently disentangled by annealing through a suitable Metropolis algorithm~\cite{true2022transitions}. In the following, we show a similar Metropolis algorithm based on the cost function $\mathcal{F}(V)\equiv (1+c(U,V))^{-1}$ to learn the mocking operator $V$. 

From the quantum machine learning point of view, we describe the procedure as follows. To learn $V$, we first train our estimator $c(U,V)$ by preparing a number $O(n^2)$ of copies of a known state. This state is used as the input of a black hole. Starting with the identity $V_0=I$, one tries the cost function $c(U,V)$ and optimizes it by local modifications using a Clifford gate $g$, so that $V_0\mapsto gV_0$. Acceptance of the gate depends on improving the cost function $c(U,V)$  or lowering it with Boltzmann probability parameterized by $\beta$, see App.~\ref{appb}. The many copies of the training journal tossed in the black hole (Fig.~\ref{figmain}) correspond to the steps of the Metropolis algorithm needed. After the training, one has settled on a $V$ that minimizes the cost function $c(U,V)$ with the number of steps allowed. One can then use $V$ in the recovery algorithm, and recover information about a new journal tossed into $U$ with fidelity $\mathcal{F}(V)$.

The numerical simulation of the whole algorithm is computationally very expensive. In order to perform the numerical simulations on a smaller space, we can exploit the fact that  we are working in the scenario $d_C\ll d_D$.  By  averaging over the Pauli group on $C$ instead of $D$, the cost function $c(U,V)$ can be computed as
\be
c(U,V)=\frac{\sum_{P_{C},P_{A}\neq 1_{A}}|\tr((U^{\dag}P_{C}VP_{A}))|^2}{\sum_{P_{C}}|\tr((U^{\dag}P_{C}V))|^2}\,,
\label{costfunction}
\ee
which can be proven using Eq.~\eqref{averagepauligroup}, see App~\ref{appb}. The simulations are conducted for a system of $n=10$ qubits initialized in the state $\ket{0}^{\otimes n}$, with $n_A =2$ and $n_C=1,2$.  $U$ is a random Clifford circuit consisting of $t$ layers each with $O(n^2)$ local gates controlled-NOT ($CNOT$), $H$, $P$ interspersed by a single $T$ gate per layer~\cite{leone2021quantum, true2022transitions}. We indeed stress that $O(n^2)$ gates from the local gate set $\{CNOT,H, P\}$ are sufficient to distill any Clifford operator~\cite{aaronson2004improved}. This fact lies behind the reason why, during the training process, we make use only of $O(n^2)$ state preparations that result in $O(n^2)$ many updates $V\mapsto gV$.

The results are shown in Fig.~\ref{fig1}. As we can see, for $t=0$ a mocking operator $V$ for a black hole modeled by a pure random Clifford circuit can be learned with perfect fidelity. As the number $t$ of non-Clifford resources increases, the fidelity for the mocking operator $V$ decreases exponentially in $t$. Notice that the fidelity also depends on the size $n_C$ of the interior of the black hole. The fact that we are not learning the operator $U$ can be checked by comparing $V$ with $U$. Indeed, this would be forbidden by the barren plateau result found in~\cite{Holmes2021barren}. 
We find that  $d^{-1} |\tr U^\dag V|^2<0.004$ for every value of $t$, including $t=0$. It is important to remark that we only try to reconstruct the mocking $V$ using merely Clifford resources, in spite of the fact that the original $U$ also contains non-Clifford resources. Although searching for $V$ within the Clifford group means we cannot reconstruct perfectly $U$, the learning algorithm fails with the addition of non-Clifford resources. For instance, for $t=6, n_A=2,n_C=1$, the fidelity obtained using only Clifford gates attains a value of $.53$ versus $.4$ when allowing the inclusion of $T$ gates. Universal resources are powerful, but pollute the algorithm~\cite{true2022transitions}.

The same recovery algorithm of~\cite{Yoshida2019disentangling} can be used to employ the recovery unitary $V$ to perform quantum teleportation between Alice and Bob.  We now show that the previous  learning of the mocking $V$ can be efficiently used to perform teleportation without having any previous knowledge of $U$.  Let $\ket{\psi_A}$ be the pure state of Alice to be teleported, and let $\ket{\Psi_{out}^{\psi_A}}$ be the output state of the black hole manipulated by the decoding protocol employed by Bob, i.e.
\be
\ket{\Psi_{out}^{\psi_A}}:=\frac{1}{\sqrt{P_{\psi}}}\Pi_{DD^{\prime}}V^{*}_{B^{\prime}A^{\prime}}U_{AB}\ket{RA}\ket{BB^{\prime}}\ket{\psi_A}\,,
\ee
where $P_{\psi}$ is a normalization.
Similarly to Eq.~\eqref{fidelityunknown}, the fidelity $\mathcal{F}_{\psi}(V)$ to teleport a state $\ket{\psi_A}$, i.e. $\mathcal{F}_{\psi}(V)\equiv \tr(\psi_A\st{\psi_{out}^{\psi_A}})$, can be computed as (see App~\ref{appa})
\be
\mathcal F_{\psi}(V)=\frac
{\braket{ \tr [\psi_A\tr_B(V^{\dag}P_{C}U)]\tr[\psi_A\tr_{B}(U^{\dag}P_{C}V)]}_{P_C}
}
{\braket{\tr[\psi_A\tr_B(V^{\dag}P_C U)\tr_B(U^{\dag}P_C V)]}_{P_C}}.
\label{fidelitystatepsia}
\ee
In the above formulas we denoted $\psi_A\equiv\st{\psi_A}$. In the inset of Fig.~\ref{fig1} we plot the value $\mathcal F_{\psi}$ as a function of the doping $t$. Again, perfect teleportation is achieved for a Clifford black hole ($t=0$) while the fidelity $\mathcal F_{\psi}(V)$ decreases as non-Clifford resources are employed. 

\subsection{Quantum Complexity} We have seen that the Clifford black hole can thus be perfectly learned - again, in the sense that one can learn the mocking operator $V$ - while  this learning becomes less and less reliable with the injection of non-Clifford resources. Why is that? This is another instance of the fact~\cite{leone2021quantum,leone2021renyi} that quantum complexity arises when scrambling (that is, efficient entanglement) is combined with non-stabilizerness, or {\em magic}~\cite{campbell2010bound,Veitch2014resource,campbell2012magic,howard2017application,leone2021renyi,oliviero2022measuring}: the resource that is at the root of quantum advantage for quantum computers and the non-simulability of generic quantum systems by classical computers~\cite{Knill1996codes,aaronson2004improved}. 

While Clifford circuits are just as efficient in scrambling as a Haar-random unitary, they do not create a complex pattern of entanglement, and the fluctuations of  entanglement  are very different~\cite{leone2021quantum}. Quantum complexity is driven by the conspiracy of entanglement and non-stabilizerness (magic), or, in other words, by the complexity of entanglement~\cite{chamon2014emergent,yang2017entanglement,oliviero2022magicising}.
In particular, the ensemble fluctuations of the OTOCs defined as $\Delta \Omega_{t}:=\aver{\Omega(U)^{2}}_{U\in\mathcal{C}_{t}}-\aver{\Omega(U)}_{U\in\mathcal{C}_t}^{2}$ behave very differently, and it is not surprising that they are governed by the eight-point OTOC, which probes more fine grained properties of scrambling~\cite{leone2021quantum}, see App~\ref{appc}:
\be
\Delta \Omega_{t}\simeq\frac{1}{d_{A}^{2}d_{D}^{2}}\left[\aver{\operatorname{otoc}_8(U)}_{U\in \mathcal{C}_t}+O(d^{-2})\right]\,.
\label{fluctuationsresults}
\ee
Using the techniques introduced in~\cite{leone2021quantum,oliviero2021transitions}, one can compute Eq.~\eqref{fluctuationsresults} and find $\Delta \Omega_{t}\simeq d_{A}^{-2}d_{D}^{-2}\left(\frac{3}{4}\right)^{t}$, which interpolates between $\Delta \Omega_t=O(d_{D}^{-2})$ and $O(d_{D}^{-2}d^{-2})$ for $t=O(1), O(n)$, respectively. The relative fluctuations for the cost function $\mathcal F^{-1}$ for $U=V$ are therefore very small, revealing a barren plateau. 
However, in searching for the mocking black hole $V$ we minimize the cost function for a $V$ that is far from the ideal $U$. The situation here is similar to the effectiveness of the success of the disentangling algorithm by Metropolis algorithm~\cite{chamon2014emergent, true2022transitions} in random circuits that are at $2-$designs. A formal proof of why this kind of algorithms does converge in polynomial time is beyond the scope of this paper.


\section{Conclusions} A black hole is a very mysterious object that crunches and scrambles information. The internal dynamics of the black hole cannot be resolved even by a quantum machine learning algorithm, so information retrieval from a black hole through Hawking radiation seems hopeless. However, we have shown that one can learn a mocking-unitary that is capable of unscrambling and decoding the Hawking radiation by a Metropolis algorithm based on the observation of out-of-time-order correlation functions. The learning is possible for black holes that can be modeled by slightly polluted Clifford circuits. Highly polluted black holes cannot be learned. This result illustrates the crossover from simpler to complex quantum behavior and demonstrates how the onset of quantum chaos is driven by the conjunction of entanglement and non-stabilizerness. In view of the results on quantum certification in~\cite{leone2022magic}, it would be interesting to show that the same intractability also shows up in the number of measurements needed to reliably measure the OTOCs in order to effectively perform information retrieval.

{\em Acknowledgments.---} The authors acknowledge support from NSF award number 2014000. The work of L.L. and S.F.E.O. was supported in part by College of Science and Mathematics Dean’s Doctoral Research Fellowship through fellowship support from Oracle, project ID R20000000025727.

\appendix
\onecolumngrid
\section{Proof of Eq.~\eqref{fidelitystatepsia}}\label{appa}
 To prove Eq. \eqref{fidelitystatepsia} we can proceed using the techniques introduced in the main text, although the setup is slightly different. First, consider the initial state (output of the black hole evolution) to be $\ket{\Phi}_{B^{\prime}CD}:=U_{AB}\ket{\psi}_{A}\ket{BB^{\prime}}$. Bob first applies the decoder unitary $V^{*}_{B^{\prime}A^{\prime}}$ on the state $\ket{\Phi}_{B^{\prime}CD}\ket{A^{\prime}R^{\prime}}$ and then project it on $\Pi_{DD^{\prime}}$. The output state reads:
\be
\ket{\Phi_{out}}=\frac{1}{\sqrt{P_{\psi}}}\Pi_{DD^{\prime}}V^{*}_{B^{\prime}A^{\prime}}\ket{\Phi}_{B^{\prime}CD}\ket{A^{\prime}R^{\prime}}=\frac{1}{\sqrt{P_{\psi}}}\figbox{.075}{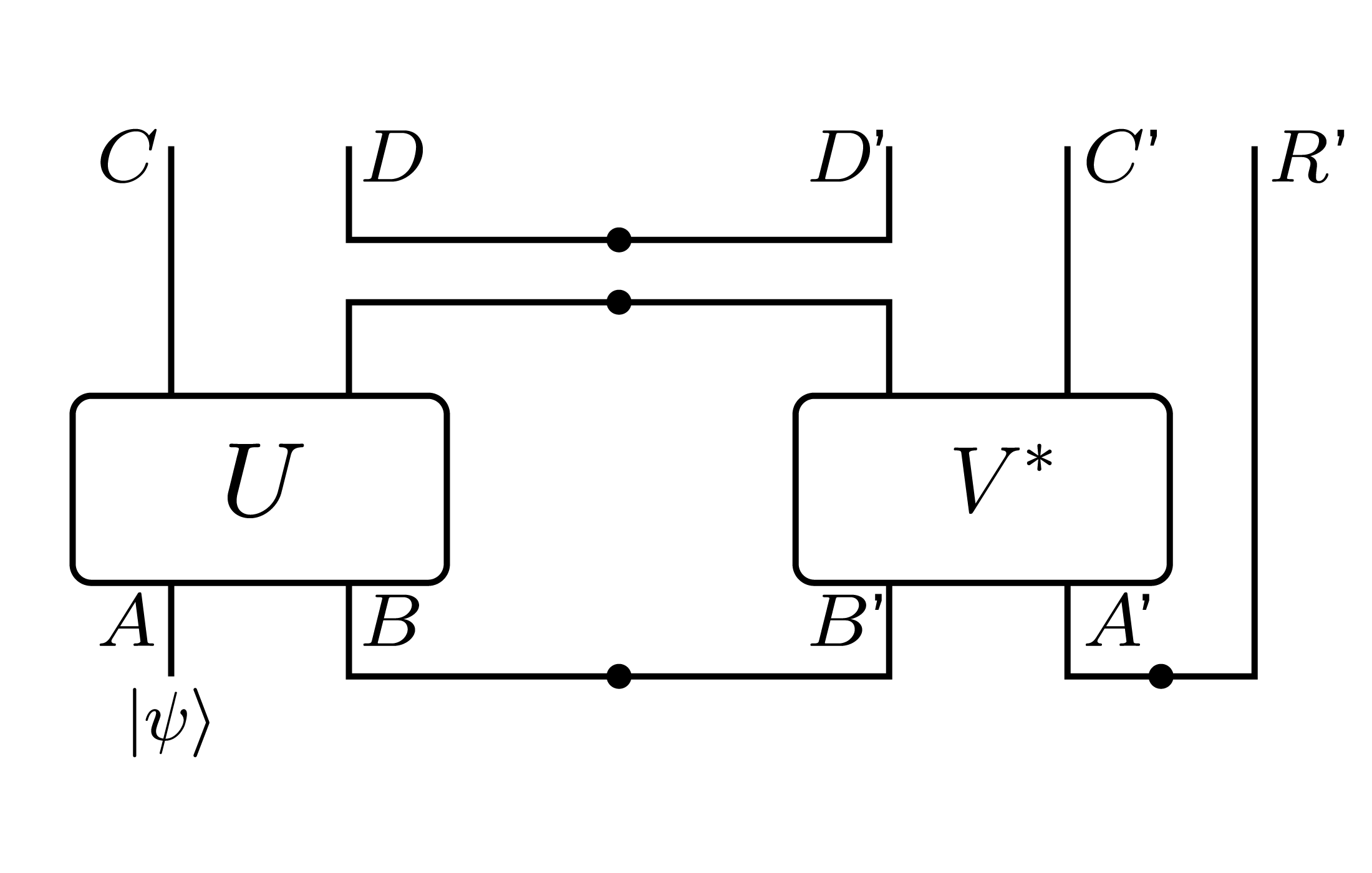}\,,
\ee
where $P_{\psi}$ is a normalization factor. Using the diagrammatical representation of $\ket{\Phi_{out}}$, one computes:
\be
\mathcal{F}_{\psi}(V)P_{\psi}=\figbox{.15}{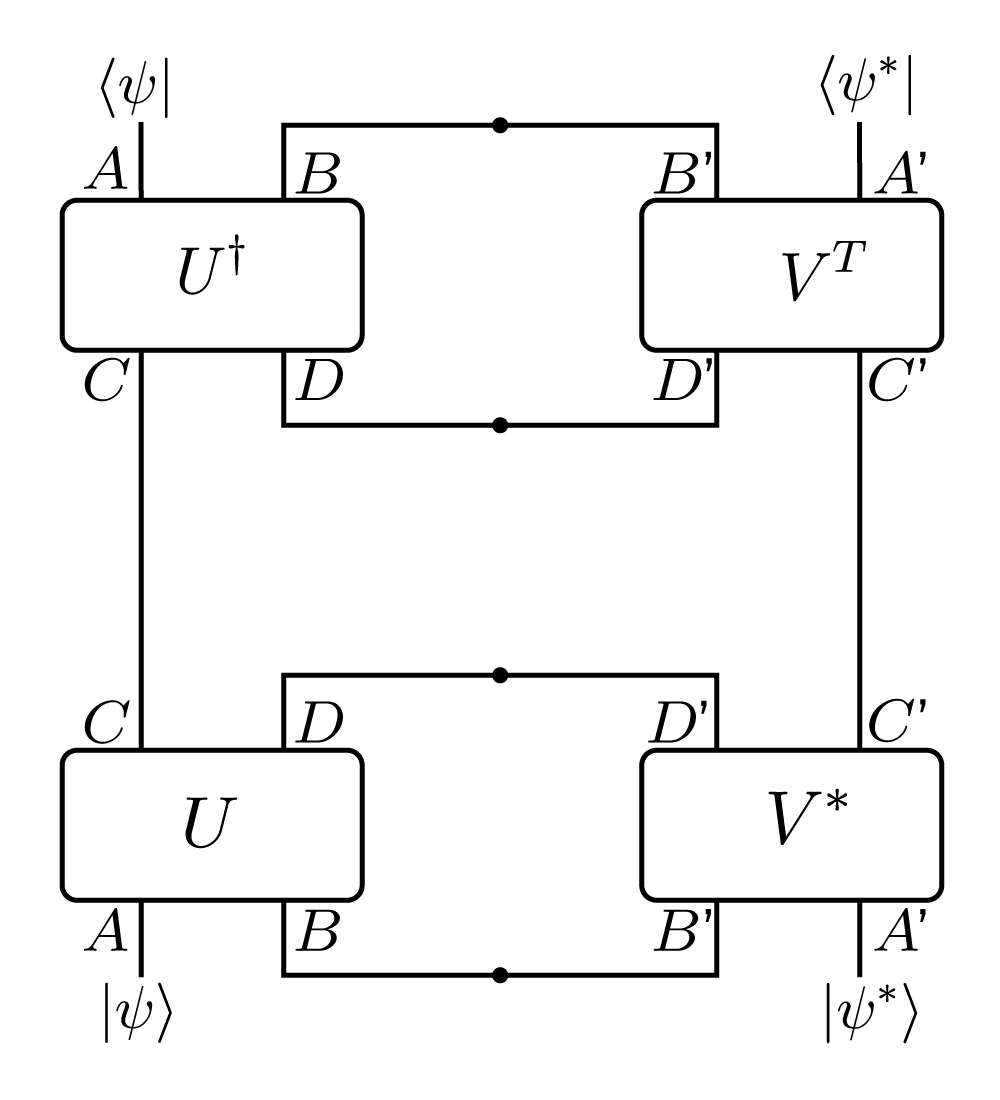}=\frac{1}{d_B}\aver{\tr [\psi_A\tr_B(V^{\dag}P_{C}U)]\tr[\psi_A\tr_{B}(U^{\dag}P_{C}V)]}_{P_C}
\label{eq1}\,,
\ee
and 
\be
P_{\psi}\equiv\figbox{.15}{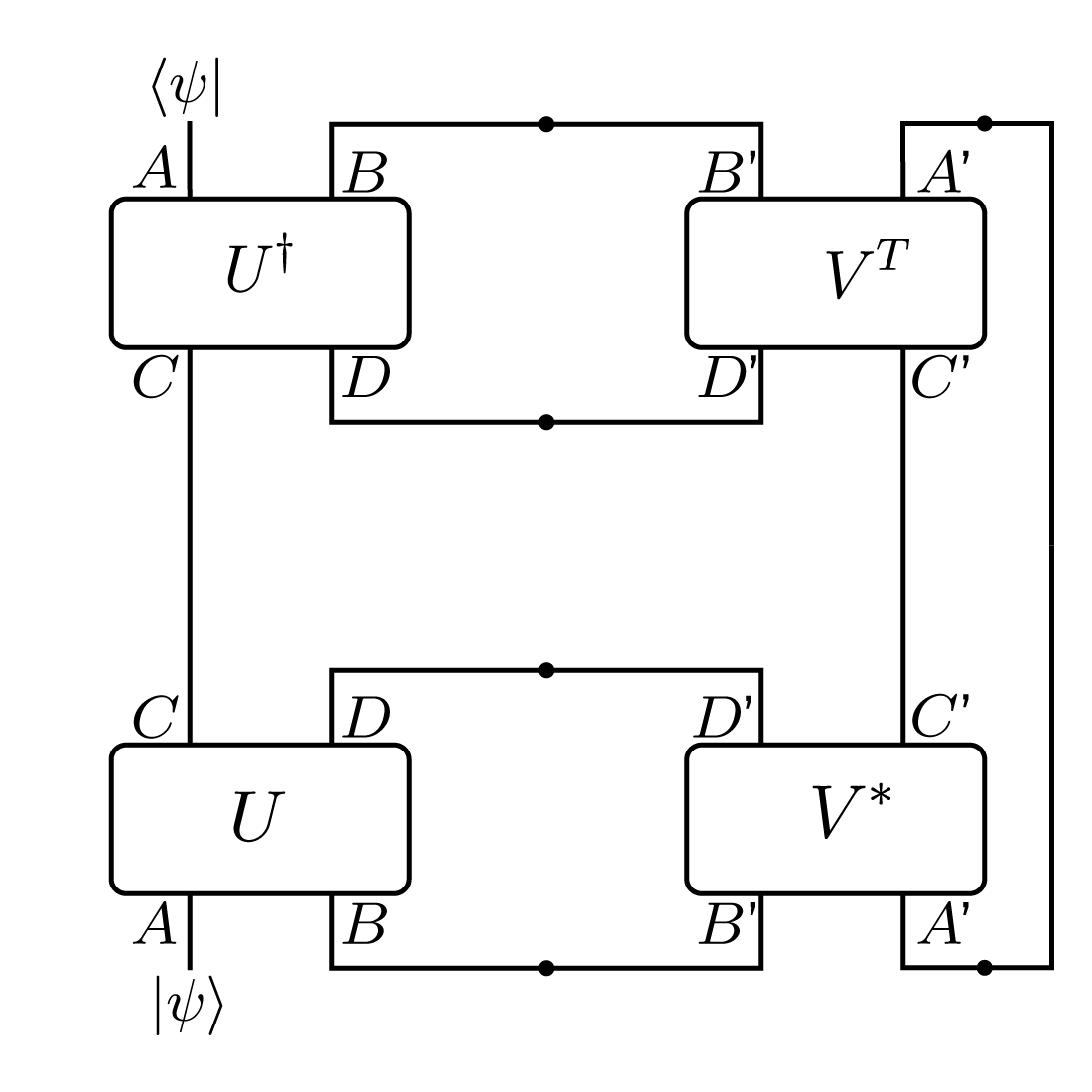}=\frac{1}{d_B}\aver{\tr[\psi_A\tr_B(V^{\dag}P_C U)\tr_B(U^{\dag}P_C V)]}_{P_C}\,,
\label{eq2}
\ee
and taking the ratio between Eq.~\eqref{eq1} and Eq.~\eqref{eq2} one obtains Eq.~\eqref{fidelitystatepsia}. 

\section{Metropolis algorithm}\label{appb}

First let us write Eq.~\eqref{fidelityunknown} in a smaller space, proving Eq.~\eqref{costfunction}. Consider the denominator of Eq.~\eqref{fidelityunknown}; using $\tr(\mathcal{O}^2)\equiv \tr(T\mathcal{O}^{\otimes 2})$ where $T$ is the swap operator defined on $\mathcal{H}^{\otimes 2}$, we can rewrite this term as:
\ba
\aver{\tr(P_{D}(U)P_{A}P_{D}(V)P_{A})}_{P_A,P_D}
&=&\aver{\tr(U^{\dag}\otimes V^{\dag}P_{D}^{\otimes2}U\otimes VP_{A}^{\otimes2}T)}_{P_D,P_A}=\frac{1}{d_Ad_D}\tr(U^{\dag}\otimes V^{\dag}T_DU\otimes VT_AT)\nonumber\\ &=&\frac{1}{d_{D}d_{A}}\tr(V^{\dag}\otimes U^{\dag}T_{C}U\otimes VT_{A} )
=\frac{1}{dd_{A}^{2}}\sum_{P_{C},P_{A}}|\tr(U^{\dag}P_CVP_A)|^2\,,
\ea
where we used the following facts: $\sum_{P_{\Lambda}}P_{\Lambda}^{\otimes 2}=d_{\Lambda}T_{\Lambda}$ where $T_{\Lambda}$ is the swap operator having support on $\Lambda$ and $TT_{\Lambda}=T_{\bar{\Lambda}}$ where $\bar{\Lambda}$ is the complement of $\Lambda$. Similarly for the numerator of Eq.~\eqref{fidelityunknown}:
\be
\aver{\tr(P_{D}(U)P_{D}(V))}_{P_D}=\frac{1}{d}\sum_{P_{C}}|\tr(U^{\dag}P_{C}V)|^2\,,
\ee
Putting it all together, we find:
\be
\mathcal{F}(V)=(1+c(U,V))^{-1}\,,
\ee
where $c(U,V)$ is defined in Eq.~\eqref{costfunction}. Note that $c(U,V)\ge 0$ with $c(U,V)=0$ if and only if $\mathcal{F}(V)=1$, making $c(U,V)$ the best candidate for a cost function. Let us describe the algorithm to numerically find an optimal recovery mocking unitary $V$:
\begin{algorithm}[H]
\caption{Training Algorithm}\label{alg1}
\algsetup{indent=2em}
\begin{algorithmic}[1]
\STATE $U\leftarrow$ random doped Clifford circuit $\mathcal{C}_{t}$
\STATE $\mathcal{F}_{max}\leftarrow\mathcal{F}(U)$
\STATE $V\leftarrow\bbbone$
\STATE $\mathcal{F}\leftarrow\mathcal{F}(V)$
\WHILE{$\mathcal{F}<\mathcal{F}_{max}$}
\STATE $V_{old} \leftarrow V$
\STATE $\mathcal{F}_{old}\leftarrow\mathcal{F}$
\STATE $g\leftarrow$ random Clifford gate
\STATE $V\leftarrow gV$
\STATE $\mathcal{F}\leftarrow\mathcal{F}(V)$
\IF{$\mathcal{F}<\mathcal{F}_{old}$}
\STATE $r\in[0, 1]$
\IF{$r<\mathrm{exp}\left[-\beta(\mathcal{F}^{-1}-\mathcal{F}_{old}^{-1})\right]$}
\STATE Undo change: $V\leftarrow V_{old}$
\STATE $\mathcal{F}\leftarrow\mathcal{F}_{old}$
\ENDIF
\ENDIF
\ENDWHILE
\end{algorithmic}
\end{algorithm}

A constant cooling schedule is employed with $\beta=250$. In each iteration of the \textit{while} loop, several $2^n \times 2^n$ matrix multiplications must be performed to obtain the operator $U P_C V^\dagger P_A$. These operations quickly become computationally expensive as we increase the system size; even the fastest algorithms for this task~\cite{Strassen1969gaussian,alman2020data} would require a computation time that grows exponentially in $n$. Furthermore, to avoid indefinite runtimes when the gradient of $c(U,V)$ becomes small, we halt the algorithm after $T_{max}$ steps. The value of $T_{max}$ used is $100n^2$ when $n_C=1$ and $300n^2$ when $n_C=2$, chosen empirically by observing the algorithm's runtime when $t=0$. Note that the time requirement grows with the number of qubits as $T_{max} \propto n^2$, as seen also in~\cite{chamon2014emergent,shaffer2014irreversibility,true2022transitions}. Therefore, the numerical analysis in this paper is performed for systems of ten qubits, as larger system sizes would require exponentially larger computational times.

The parameters for the results fit to $a\,\mathrm{exp}(-\alpha t)+b$ shown in Fig.~$3$ are given below.
{\renewcommand{\arraystretch}{2}
\begin{table}[H]
  \begin{center}
    {\setlength{\tabcolsep}{1em}
    \begin{tabular}{c|c|c|c}
      \multicolumn{2}{c|}{} & $n_C=1$ & $n_C=2$\\
      \hline
      \multicolumn{2}{c|}{\bm{$\alpha$}} & $0.167$ & $0.129$\\
      \hline
      \multirow{2}{*}{\bm{$\overline{\mathcal{F}_f}}^{(U,V)}$} & \bm{$a$} & $0.7243$ & $0.8483$\\
      \cline{2-4}
      & \bm{$b$} & $0.2757$ & $0.1517$\\
      \hline
      \multirow{2}{*}{\bm{$\overline{\mathcal{F}_{\psi}}}^{(U,V)}$} & \bm{$a$} & $0.5860$ & $0.6794$\\
      \cline{2-4}
      & \bm{$b$} & $0.4140$ & $0.3206$\\
      \hline
    \end{tabular}}
    \label{table1}
  \end{center}
\end{table}}

\section{The complexity of scrambling: the EIGHT-POINT OTOC}\label{appc}
In this section, we present a single correlation function that tells us whether the algorithm is going to fail or not, i.e. the eight-point out of time order correlator, proving Eq.~\eqref{fluctuationsresults}. First, let us write the OTOC $\Omega(U)=\aver{\tr(P_{A}P_{D}(U)P_AP_{D}(U))}_{P_A,P_D}$ as:
\be
\Omega(U)=\frac{1}{d_{A}^{2}}+\frac{1}{d_{D}^{2}}-\frac{1}{d_{A}^{2}d_{D}^{2}}+\frac{1}{d_{A}^{2}d_{D}^{2}}\sum_{P_{A},P_{D}\neq \bbbone}\frac{1}{d}\tr(P_{A}P_{D}(U)P_{A}P_{D}(U))
\ee
and we thus can alternately define a scrambling unitary $U$ as one such that:
\be
f(U)\equiv \frac{1}{d_{A}^{2}d_{D}^{2}}\sum_{P_{A},P_{D}\neq \bbbone}\operatorname{otoc}_4(P_{A},P_{D}(U))=O(d^{-1})\,,
\label{alternscramblingdef}
\ee
where we defined $\operatorname{otoc}_4(P_{A},P_{D}(U)):=\frac{1}{d}\tr(P_{A}P_{D}(U)P_{A}P_{D}(U))$. Note that for $U\in\mathcal{C}(2^n)$ being a Clifford operator $\operatorname{otoc}_4(P_{A},P_{D}(U))=\pm 1$ for any choice of $P_A,P_D$, but $f(U)=O(d^{-1})$ still holds for Clifford unitaries, i.e. Clifford unitaries are scramblers. Note also that 
\be
\aver{f(U)}_U=\frac{(d_{A}^2-1)(d_{D}^{2}-1)}{d_{A}^{2}d_{D}^{2}}\aver{\operatorname{otoc}_4(P_{1},P_{2}(U))}_U\,,
\ee
for any $P_{1},P_{2}$ chosen to be two non-identity Pauli operators and $U$ being a Clifford or $t-$doped Clifford operator. Now let us look at the fluctuations in the ensemble of $t-$doped Clifford operators $U\in\mathcal{C}_{t}$. Direct calculation leads to:
\be
\Delta\Omega_t=\aver{f(U)^{2}}_{U}-\aver{f(U)}_{U}^{2}=\aver{f(U)^{2}}_{U}+O(d^{-2})\,,
\ee
and the second equality follows from the definition of scrambling unitaries -- i.e. $\aver{f(U)}_{U}=O(d^{-1})$ see Eq.~\eqref{alternscramblingdef} -- and it holds for any $t\in\mathbb{N}$. Thus, turn to analyze $\aver{f(U)^{2}}_{U}$:
\be
\aver{f(U)^{2}}_{U}=\frac{(d_{A}^{2}-1)(d_{D}^{2}-1)}{d_{A}^{4}d_{D}^{4}}\aver{\operatorname{otoc}_4(P_{1},P_{2}(U))^{2}}_{U}+\aver{R(U)}_{U}\,,
\ee
\be
R(U):=\frac{1}{d_{A}^{4}d_{D}^{4}}\sum_{P_{A},P_{D},P_{A}^{\prime},P_{D}^{\prime}\neq \bbbone}[1-\delta(P_{A}=P^{\prime}_{A})\delta(P_{D}=P^{\prime}_{D})]\times \operatorname{otoc}_4(P_{A},P_{D}(U))\operatorname{otoc}_4(P^{\prime}_{A},P^{\prime}_{D}(U))\,.
\ee
Using standard techniques of the Haar measure over groups~\cite{collins2003moments,collins2006integration,leone2021quantum,Oliviero2020random} one proves that for any $t\in\mathbb{N}$ we have $\aver{R(U)}=O(d^{-2})$. Lastly, using the fact that the Pauli group forms a $1$-design~\cite{roberts2017chaos}, it is easy to prove~\cite{roberts2017chaos} the following:
\be
\operatorname{otoc}_4(P_1,P_{2}(U))^{2}=\operatorname{otoc}_{8}(U)\,,
\ee
where we defined the following eight-point out of time order correlation function:
\be
\operatorname{otoc}_{8}(U):=\frac{1}{d}\aver{\tr(P_{1}P_{2}(U)P_{1}P_{2}(U)PP_{1}P_{2}(U)P_{1}P_{2}(U)P)}_{P\in\mathcal{P}_n}\,,
\label{8otocdef}
\ee
where $P_{1}$ and $P_{2}$ are any non identity Pauli operators. We can finally write the following formula:
\be
\Delta\Omega_t=\frac{(d_{A}^{2}-1)(d_{D}^{2}-1)}{d_{A}^{4}d_{D}^{4}}\aver{\operatorname{otoc}_{8}(U)}_{U}+O(d^{-2})\,,
\ee
and in the limit of $d_{D}\gg d_{A} \gg 1$, one arrives to Eq.~\eqref{fluctuationsresults}. Thus the ensemble fluctuations of the four-point OTOC are proportional to the ensemble the average of an eight-point correlation function, which probes more fine-grained properties of scrambling, i.e. the complexity of scrambling. We thus define the complexity of scrambling produced by some unitary operator $U$ as the behavior of the eight-point correlation function defined in Eq.~\eqref{8otocdef}.




%

\clearpage
\onecolumngrid
\end{document}